%% file: main.tex
\documentclass{article}
\def\publicrelease{}

\usepackage[utf8]{inputenc}
\usepackage{graphicx}
\usepackage{amsmath}
\usepackage{booktabs}
\usepackage{hyperref}
\usepackage[margin=1in]{geometry}

\title{Inference-Sufficient Representations for High-Throughput Measurement: Lessons from Lossless Compression Benchmarks in 4D-STEM}

\author{
    Ondrej Dyck\textsuperscript{1,*} \\
    Andrew R. Lupini\textsuperscript{1} \\
    Albina Borisevich\textsuperscript{1} \\
    Miaofang Chi\textsuperscript{1} \\
    Rama K. Vasudevan\textsuperscript{1} \\
    Stephen Jesse\textsuperscript{1} \\
}

\date{}

\newcommand{\affiliation}{
    \textsuperscript{1}Center for Nanophase Materials Sciences, Oak Ridge National Laboratory, Oak Ridge, TN 37831, USA \\
    \textsuperscript{*}Corresponding author: dyckoe@ornl.gov
}

\begin{document}

\ifdefined\publicrelease
\else
\begin{titlepage}
\thispagestyle{empty}
\vspace*{2cm}
\noindent\textbf{Notice to the editor (not to be published):} This manuscript has been authored by UT-Battelle, LLC, under Contract No. DE-AC05-00OR22725 with the U.S. Department of Energy. The United States Government retains and the publisher, by accepting the article for publication, acknowledges that the United States Government retains a non-exclusive, paid-up, irrevocable, world-wide license to publish or reproduce the published form of this manuscript, or allow others to do so, for United States Government purposes. The Department of Energy will provide public access to these results of federally sponsored research in accordance with the DOE Public Access Plan (\url{http://energy.gov/downloads/doe-public-access-plan}).
\end{titlepage}
\fi
\setcounter{page}{1}

\maketitle

\affiliation

\begin{abstract}
Four-dimensional scanning transmission electron microscopy (4D-STEM) generates multi-gigabyte datasets, creating a growing mismatch between acquisition rates and practical storage, transfer, and interactive visualization capabilities. We systematically benchmark 13 lossless compression implementations across 5 representative datasets (8~MiB to 8~GiB, 49.5--92.8\% sparsity), with 10 independent runs per method. HDF5 provides built-in gzip compression, of which gzip-9 typically achieves the highest compression ratio but is slow. We therefore evaluate widely available alternatives (via hdf5plugin), including the Blosc family. As a representative comparison, blosc\_zstd achieves compression comparable to gzip-9 (mean 13.5$\times$ vs 12.3$\times$) while compressing 19--69$\times$ faster and reading 1.9--2.6$\times$ faster across datasets. Compression ratios are deterministic, and timing measurements are highly reproducible (CV $<$2\%). Compression performance follows a power law with sparsity ($R^2 = 0.99$), ranging from 5$\times$ for moderately sparse data to 35$\times$ for highly sparse data. We identify six top-performing implementations optimized for different use cases and demonstrate that 4D-STEM data can be routinely compressed by $>$10$\times$. While these results provide practical guidance for lossless compression selection, the broader conclusion is that lossless compression preserves measurements but does not by itself guarantee sustainable high-throughput workflows. As detector rates rise, data handling will increasingly require inference-driven representations---i.e., deciding what must be preserved to support a scientific inference, rather than defaulting to storing fully dense raw measurements.
\end{abstract}

\section{Introduction}
\input{sections/introduction}

\section{Methods}
\input{sections/methods}

\section{Results}
\input{sections/results}

\section{Discussion}
\input{sections/discussion}

\section{Conclusions}
\input{sections/conclusions}

\section*{Acknowledgments}
This work was supported by the U.S. Department of Energy, Office of Science, Basic Energy Sciences, Materials Sciences and Engineering Division and by the Oak Ridge National Laboratory's Center for Nanophase Materials Sciences (CNMS), a U.S. Department of Energy, Office of Science User Facility. AYB was sponsored by the Laboratory Directed Research and Development Program of Oak Ridge National Laboratory, managed by UT-Battelle, LLC, for the US Department of Energy.

\bibliographystyle{unsrt}
\bibliography{references}

\end{document}

%% file: sections/introduction.tex

Four-dimensional scanning transmission electron microscopy (4D-STEM) has emerged as a powerful family of techniques for characterizing materials by recording a diffraction pattern at each probe position in a 2D scan, enabling downstream reconstruction of virtual imaging modes and quantitative analysis beyond conventional STEM signals \cite{ophusFourdimensionalScanningTransmission2019}. The resulting datasets are information-rich but also large, and they increasingly expose a mismatch between acquisition bandwidth and practical resources for storage, transfer, and interactive analysis. Importantly, this mismatch is not unique to any one modality: as detector performance improves across electron and photon sciences, the rate at which experiments can generate data is growing faster than the infrastructure typically available to move and retain it. Recent detector and acquisition system designs explicitly target very high sustained data rates and motivate specialized pipelines to ingest and process streams ranging from tens of Gb\,s$^{-1}$ to tens of GB\,s$^{-1}$ \cite{llopartTimepix4LargeArea2022,correaTEMPUSTimepix4basedSystem2024,leonarskiJungfraujochHardwareacceleratedDataacquisition2023,DECTRISARINAHybridpixel,DirectElectronDetection}.

In this work we use 4D diffraction and momentum-resolved EELS datasets as concrete, representative examples of this broader data-management problem. In what follows, we use ``4D-EELS'' as a shorthand for momentum-resolved EELS acquired over a 2D scan, producing datasets with two spatial dimensions plus energy and momentum (scattering-angle) dimensions. Such datasets can exceed the size of comparably sampled 4D diffraction datasets because the energy-dispersion axis demands a larger detector extent along the spectral dimension. For example, while a typical 4D diffraction dataset may record 256$\times$256 detector pixels per scan position, our Merlin detector in EELS mode records 256$\times$1024 pixels per scan position, increasing the per-position data volume by a factor of four. These examples illustrate a general point: as experimentalists push toward higher throughput, higher dynamic range, and higher resolution, data volume becomes a first-order experimental constraint rather than a downstream nuisance.

Lossless compression provides an immediate and widely applicable baseline for reducing this burden. In the HDF5 ecosystem, the gzip family is broadly available and often serves as a default; gzip-9 typically provides the strongest compression among the built-in options but can be prohibitively slow for high-throughput workflows. This motivates a practical question: are there drop-in compression implementations that provide comparable compression ratios while substantially improving write and read performance, without changing file formats or compromising numerical fidelity?

Here we present a systematic benchmark of 13 readily available, lossless compression implementations for representative 4D-STEM datasets. We quantify compression ratio, write throughput, read throughput, and reproducibility across 10 independent runs, spanning five datasets (8\,MiB to 8\,GiB) and multiple chunking strategies. Our goal is pragmatic: to provide implementation-level guidance for selecting compression filters that improve storage efficiency and I/O performance in common Python/HDF5 workflows.

Beyond implementation-level performance, this benchmark highlights a broader implication for high-throughput microscopy: lossless compression is valuable, but it is not a complete solution. As detector throughput continues to rise, lossless compression can reduce---but cannot eliminate---the mismatch between acquisition rates and practical storage and I/O budgets. In throughput-limited regimes, sustainable workflows will increasingly depend on inference-driven data reduction and interpretation pipelines that store task-relevant reduced representations rather than fully dense raw measurements. Event-based representations provide one concrete example of this principle in high-rate microscopy workflows \cite{pelzRealTimeInteractive4DSTEM2022}.

%% file: sections/methods.tex

\subsection{Datasets}

We evaluated compression performance on five representative datasets spanning different acquisition modes, detector configurations, and sparsity levels (Table~\ref{tab:datasets}).

\input{Tables/table_datasets.tex}

All datasets were acquired using a Quantum Detectors MerlinEELS detector \cite{ElectronEnergyLoss} and initially stored in the native \texttt{.mib} format \cite{krajnakMatkrajRead_mib2021} before conversion to EMD 1.0 (HDF5-based) \cite{savitzkyEMD10Emdfile2023}. The datasets span a $\sim$10$^3$-fold range in size (8.0~MiB to 8.0~GiB) and represent typical experimental conditions. The high sparsity (49.5--92.8\% zeros) is characteristic of electron diffraction and spectroscopy data, where most detector pixels record no signal. The raw 4D datasets (4D\_Diff and 4D\_EELS) are stored as \texttt{uint16} and, for these acquisitions, exhibit an effective dynamic range of 12 bits (maximum observed value 4095). The binned diffraction datasets were generated by averaging and stored as \texttt{float32}. The 3D\_EELS dataset is a y-summed (integrated) spectrum image and therefore makes fuller use of the \texttt{uint16} container range. The binned datasets enable evaluation of how detector binning affects both file size and compression performance.

\subsection{Compression Implementations}

We evaluated 13 lossless compression implementations spanning four categories: HDF5 built-in methods, advanced compression via the hdf5plugin library, sparse matrix storage, and custom strategies (Table~\ref{tab:algorithms}). All methods are available in standard Python scientific computing libraries, ensuring reproducibility and ease of adoption.

\begin{table}[h]
\centering
\caption{Compression implementations evaluated in this study}
\label{tab:algorithms}
\small
\begin{tabular}{llp{6cm}l}
\hline
\textbf{Algorithm} & \textbf{Variant} & \textbf{Characteristics \& Use Case} & \textbf{Ref.} \\
\hline
\multicolumn{4}{l}{\textit{HDF5 Built-in Compression}} \\
\hline
gzip & level 1 & Fast DEFLATE; minimal overhead & \cite{deutschDEFLATECompressedData1996} \\
     & level 6 & Balanced; general-purpose default & \cite{deutschDEFLATECompressedData1996} \\
     & level 9 & Maximum compression; archival & \cite{deutschDEFLATECompressedData1996} \\
LZF & - & Fast HDF5-optimized; interactive use & \cite{folkOverviewHDF5Technology2011} \\
szip & - & Extended-Rice algorithm; scientific data & \cite{Documentation} \\
\hline
\multicolumn{4}{l}{\textit{Advanced Compression (Blosc)}} \\
\hline
Blosc & blosclz & Native codec; byte-shuffling & \cite{altedWhyModernCPUs2010} \\
      & lz4 & Ultra-fast; real-time applications & \cite{colletZstandardCompressionApplication2018} \\
      & lz4hc & High-compression LZ4 variant & \cite{colletZstandardCompressionApplication2018} \\
      & zlib & Standard zlib in Blosc framework & \cite{deutschDEFLATECompressedData1996} \\
      & zstd & Modern; excellent balance & \cite{colletZstandardCompressionApplication2018} \\
\hline
\multicolumn{4}{l}{\textit{Other Advanced Methods}} \\
\hline
LZ4 & standalone & Extremely fast decompression & \cite{colletZstandardCompressionApplication2018} \\
Bitshuffle+LZ4 & - & Bit-shuffling; sparse data & \cite{masuiCompressionSchemeRadio2015} \\
\hline
\multicolumn{4}{l}{\textit{Sparse Matrix Storage}} \\
\hline
CSR & - & Compressed Sparse Row; $>$95\% sparse & \cite{saadIterativeMethodsSparse2003} \\
\hline
\multicolumn{4}{l}{\textit{Custom Strategies}} \\
\hline
uint8+overflow & - & 8-bit downcast; low counts & - \\
Simple gzip & - & Direct array compression & \cite{deutschDEFLATECompressedData1996} \\
\hline
\end{tabular}
\end{table}

HDF5 built-in methods (gzip, LZF, szip) provide baseline performance and are universally supported in the HDF5 ecosystem. Advanced methods via hdf5plugin include the Blosc meta-compressor framework with five codec variants (blosclz, lz4, lz4hc, zlib, zstd), standalone LZ4 optimized for extremely fast decompression, and Bitshuffle which performs bit-level shuffling before compression to improve performance on numerical data. All Blosc-provided implementations use byte-shuffling and blocking to reorder data for better compression of numerical arrays.

Sparse matrix storage uses SciPy's Compressed Sparse Row (CSR) format, storing each diffraction pattern as a sparse matrix. This approach is particularly suitable for highly sparse datasets ($>$95\% zeros) but requires custom loading code and does not integrate with standard HDF5 workflows.

When considering data reduction strategies a natural question that arises is whether efficiency may be found by simply reducing the bit depth of the data. A custom strategy we attempted includes uint8 downcasting with overflow handling, which exploits the fact that most detector pixels have low counts ($<$255) and can be stored in 8-bit format with a separate overflow map for rare high-count pixels. We find that this strategy is significantly worse than standard compression implementations. Simple gzip compression of numpy arrays provides a baseline comparison without HDF5 overhead.

\subsection{Chunking Strategies}
\label{sec:chunking}

HDF5 chunking divides the 4D dataset into smaller blocks that are compressed independently. Chunk size affects both compression ratio and access performance. We tested three chunking strategies optimized for different access patterns. Real-space optimized chunks (32, 32, full detector) favor accessing contiguous scan regions for virtual imaging and spatial feature analysis. Balanced chunks (16, 16, half detector) provide a compromise between scan-space and detector-space access. Single-frame chunks (1, 1, full detector) optimize for accessing individual diffraction patterns. The detector dimensions in each chunk size scale with the actual detector configuration (e.g., 256$\times$256 for unbinned data, 128$\times$128 for 2$\times$2 binned data), ensuring valid HDF5 configurations across all datasets. Each HDF5 compression implementation was tested with all three chunking strategies, yielding 39 HDF5 method combinations plus sparse and custom methods.

\subsection{Benchmark Methodology}

\subsubsection{Performance Metrics}

For each implementation and dataset combination, we measured compression ratio (original size divided by compressed size), write time (time to compress and save the entire 4D dataset), read time (time to load and decompress the entire 4D dataset), and file size (actual disk space used including HDF5 metadata overhead). Write throughput was calculated from the original uncompressed data size divided by write time. Read throughput was calculated from the decompressed data size (bytes read into memory) divided by read time. Compression ratio directly impacts storage costs and network transfer times. Write time represents the one-time preprocessing cost; while write times of several minutes are acceptable for archival storage, faster compression is preferred for interactive workflows. Read time is critical for data analysis workflows where the entire dataset must be loaded into memory. File size determines storage infrastructure requirements.

\subsubsection{Reproducibility Assessment}

To assess measurement reproducibility, each compression implementation was benchmarked 10 times on each dataset. Results were aggregated to compute mean, standard deviation, minimum, maximum, median, and coefficient of variation (CV\%) for all metrics. This multi-run approach quantifies both the deterministic nature of compression algorithms (compression ratio should be identical across runs) and the variability in timing measurements due to system-level factors such as CPU scheduling, cache state, and background processes.

All benchmarks were performed on a workstation with Intel Xeon processor, 64 GB RAM, and SSD storage running Linux. The absolute write/read times depend on system specifications; therefore, meaningful comparisons are relative performance differences between implementations on the same system rather than absolute timing values. The complete benchmark suite (13 implementations $\times$ 3 chunking strategies $\times$ 5 datasets = 195 configurations) was executed 10 times over a 24-hour period, yielding 2,100 individual measurements. The system was otherwise idle during benchmarking to ensure consistent performance measurements.

\subsection{Implementation}

\textit{Software Architecture}---The benchmark was implemented in Python 3.11 using the following standard scientific computing libraries:

\begin{itemize}
    \item NumPy (v2.3): Array operations and data manipulation
    \item h5py (v3.15): HDF5 file I/O
    \item SciPy (v1.16): Sparse matrix storage
    \item hdf5plugin (v6.0): Advanced compression filters
    \item matplotlib (v3.10): Visualization and plotting
\end{itemize}

\textit{Benchmark Workflow}---The benchmark followed a systematic workflow for each compression implementation. First, the 4D dataset was loaded from EMD 1.0 format and analyzed for sparsity (zero fraction, value distribution, and statistics). For each implementation and chunking strategy combination, an HDF5 file was created with the specified compression and chunk size, the entire 4D dataset was compressed and written while recording write time and compressed file size. Read performance was then tested by loading the complete dataset to measure decompression speed. Compression ratios were calculated and results tabulated for further analysis. This entire process was repeated 10 times for each configuration to assess reproducibility.

\textit{Data Format}---Input data uses the EMD 1.0 (Electron Microscopy Dataset) format, an HDF5-based standard with the structure \texttt{/version\_1/data/datacubes/datacube\_000/data}. This ensures compatibility with existing 4D-STEM analysis tools including py4DSTEM \cite{savitzkyPy4DSTEMSoftwarePackage2021} and LiberTEM \cite{clausenLiberTEMSoftwarePlatform2020}.

\textit{Reproducibility}---Source code and detailed results are available at: \url{https://github.com/ondrejdyck/4d-stem-compression-benchmark-public}.

Claude and ChatGPT were used for coding assistance and language editing; all scientific analysis, interpretation, and final manuscript content were reviewed by the authors.

%% file: Tables/table_datasets.tex
\begin{table}[h]
\centering
\caption{Datasets used for compression benchmarking}
\label{tab:datasets}
\begin{tabular}{lllrl}
\hline
\textbf{Dataset} & \textbf{Shape} & \textbf{Size} & \textbf{Sparsity} & \textbf{Description} \\
\hline
4D\_EELS & (64, 64, 256, 1024) & 2.0 GiB & 92.8\% & Full 4D EELS spectrum imaging \\
4D\_Diff & (256, 256, 256, 256) & 8.0 GiB & 74.7\% & 4D STEM diffraction (unbinned) \\
4D\_Diff-2x2 & (256, 256, 128, 128) & 4.0 GiB & 70.9\% & 4D STEM diffraction (2$\times$2 binned) \\
4D\_Diff-4x4 & (256, 256, 64, 64) & 1.0 GiB & 60.9\% & 4D STEM diffraction (4$\times$4 binned) \\
3D\_EELS & (64, 64, 1, 1024) & 8.0 MiB & 49.5\% & Y-summed EELS spectrum image \\
\hline
\end{tabular}
\end{table}

%% file: sections/results.tex

\subsection{Overview of Compression Performance}

We evaluated 13 compression implementations across 5 datasets ranging from 8~MiB to 8~GiB, with sparsity levels from 49.5\% to 92.8\%. Each implementation was tested with three chunking strategies (real-space, balanced, and single-frame), yielding 39 HDF5-based configurations per dataset plus sparse and custom methods. Table~\ref{tab:dataset_summary} summarizes the key characteristics of each dataset and the best compression achieved.

Throughout the Results, we treat these lossless, drop-in implementations as a practical baseline for file-based workflows; broader implications for throughput-driven data reduction are discussed in the Discussion.

Compression ratios ranged from 4.8$\times$ for the least sparse dataset (3D\_EELS, 49.5\% zeros) to 34.9$\times$ for the most sparse dataset (4D\_EELS, 92.8\% zeros). The median compression ratio across all implementations and datasets was 6.2$\times$, demonstrating that 4D STEM data is highly compressible regardless of the specific implementation chosen.

\input{Tables/table_dataset_summary.tex}

\begin{figure}[htbp]
\centering
\includegraphics[width=0.8\textwidth]{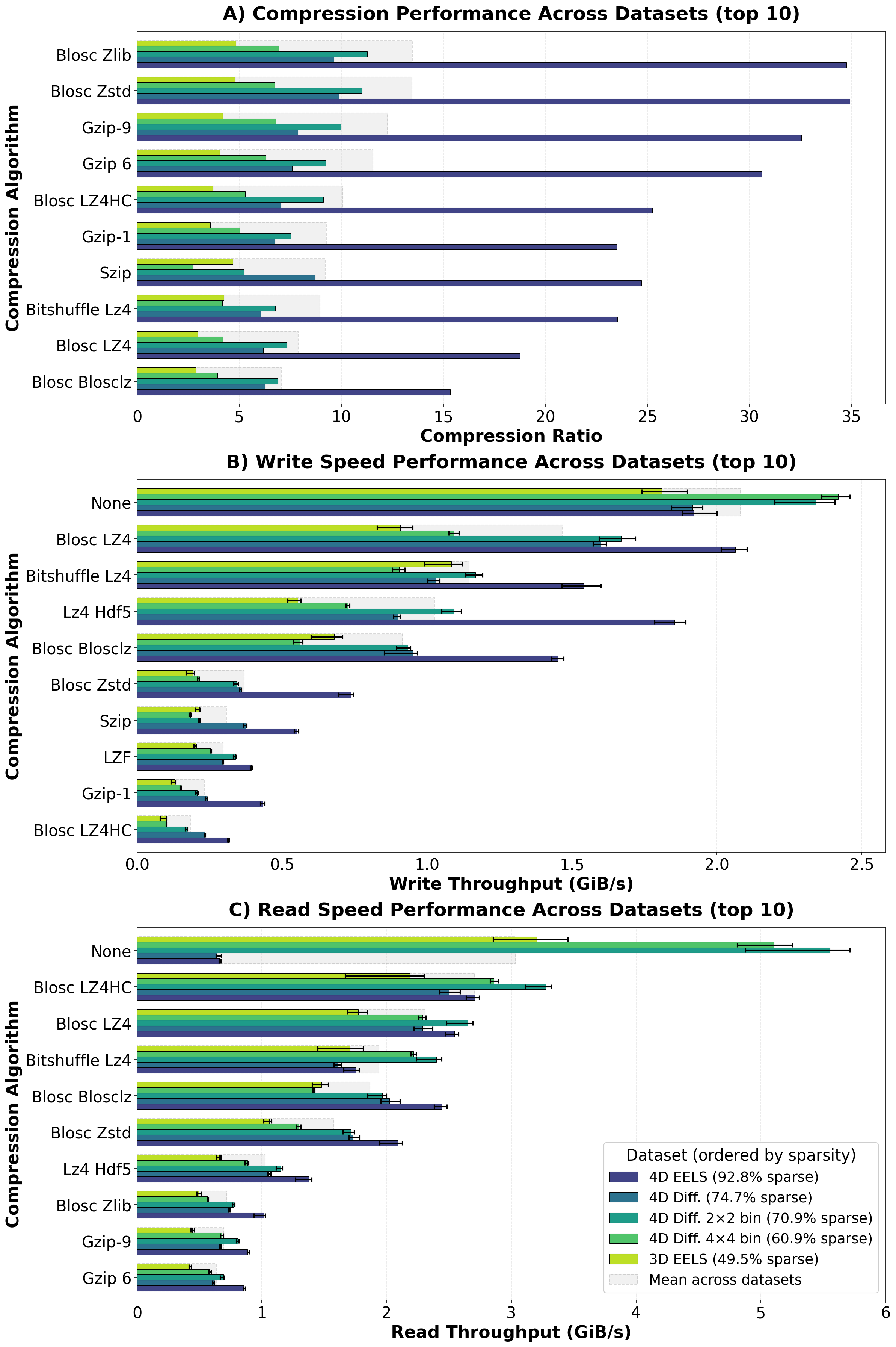}
\caption{Cross-dataset performance comparison of top 10 compression implementations. (A) Compression ratios showing blosc\_zstd and blosc\_zlib achieving the highest compression across all datasets. Panel A has no error bars as compression ratio is deterministic (0\% variation across 10 runs). (B) Write throughput demonstrating that Blosc-based implementations are significantly faster than traditional gzip compression. (C) Read throughput showing that Blosc implementations enable high-speed data access from compressed files. Error bars on panels B and C show minimum and maximum values across 10 benchmark runs. Results averaged across all five datasets.}
\label{fig:cross_dataset_performance}
\end{figure}

Figure~\ref{fig:cross_dataset_performance} shows the compression ratios, write throughput, and read throughput achieved by the top-performing implementations across all datasets. Blosc-based implementations consistently outperformed traditional HDF5 compression filters, with blosc\_zstd and blosc\_zlib achieving the highest compression ratios in all cases.

\subsection{Algorithm Family Comparison}

\begin{figure}[htbp]
\centering
\includegraphics[width=0.5\textwidth]{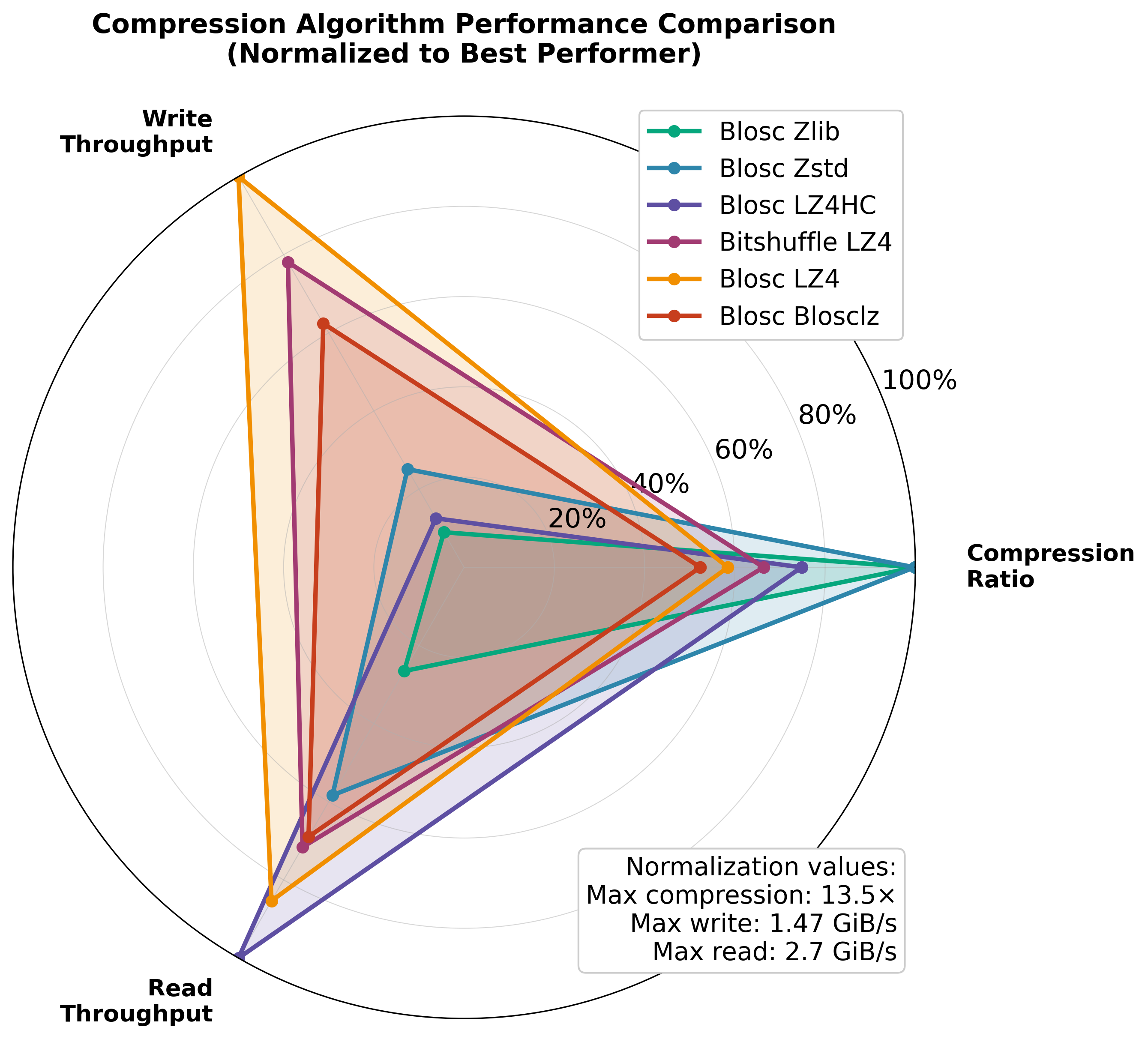}
\caption{Multi-dimensional performance comparison of representative compression implementations. The radar chart shows normalized performance across three metrics: compression ratio, write throughput, and read throughput. Six key implementations are shown: five that appear in the top 10 for all three metrics across all datasets, plus blosc\_zlib which achieves the highest compression ratio despite slower write speed. Blosc-based implementations (Blosc Zstd, Blosc Zlib, Blosc LZ4, Blosc LZ4HC) achieve well-balanced performance across all dimensions, while traditional implementations like Gzip-9 excel in compression but lag in speed. The ideal implementation would reach the outer edge in all three dimensions. Based on aggregated statistics from 10 benchmark runs.}
\label{fig:algorithm_radar}
\end{figure}

Blosc-based compression implementations demonstrated superior performance compared to traditional HDF5 built-in filters across all datasets. Table~\ref{tab:algorithm_families} summarizes the mean performance of each implementation family using balanced chunking for fair comparison, while Figure~\ref{fig:algorithm_radar} provides a multi-dimensional comparison of the top-performing implementations.

\input{Tables/table_implementation_families.tex}

Blosc-based implementations achieve compression ratios comparable to or better than gzip-9 while providing substantially higher write throughput, but it is useful to distinguish two different comparisons. First, for a like-for-like baseline focused on maintaining strong compression, blosc\_zstd achieves compression comparable to gzip-9 (13.5$\times$ vs 12.3$\times$ mean for balanced chunking) while writing 19--69$\times$ faster across datasets and reading 1.9--2.6$\times$ faster. Second, for throughput-limited scenarios where sustained write performance is the dominant constraint, blosc\_lz4 provides the fastest writing among the tested families, at the expense of lower compression. Relative to gzip-9, blosc\_lz4 reduces write time by a factor of 87--324$\times$ across datasets (balanced chunking); summarizing the per-dataset speedups yields a typical improvement of $\sim$175$\times$ (geometric mean). This throughput-first comparison is relevant for high-rate acquisition pipelines, where write performance directly impacts buffering requirements and the feasibility of near-real-time ingestion, even though software-only compression is unlikely to match the peak output rates of modern detectors.

The superior performance of Blosc implementations can be attributed to their use of byte-shuffling, blocking, and CPU vectorization~\cite{altedWhyModernCPUs2010}, which are particularly effective for numerical scientific data. Traditional compression algorithms like gzip were designed for general-purpose text compression and do not exploit the structure of numerical arrays.

Read performance also favored Blosc implementations. For the 8~GiB dataset, blosc\_zstd achieved 1.7~GiB/s read throughput compared to 0.7~GiB/s for gzip-9, providing 2.6$\times$ faster decompression. This enables efficient data loading for analysis workflows where the entire dataset must be read into memory.

\subsection{Effect of Data Sparsity on Compression}

\begin{figure}[htbp]
\centering
\includegraphics[width=0.5\textwidth]{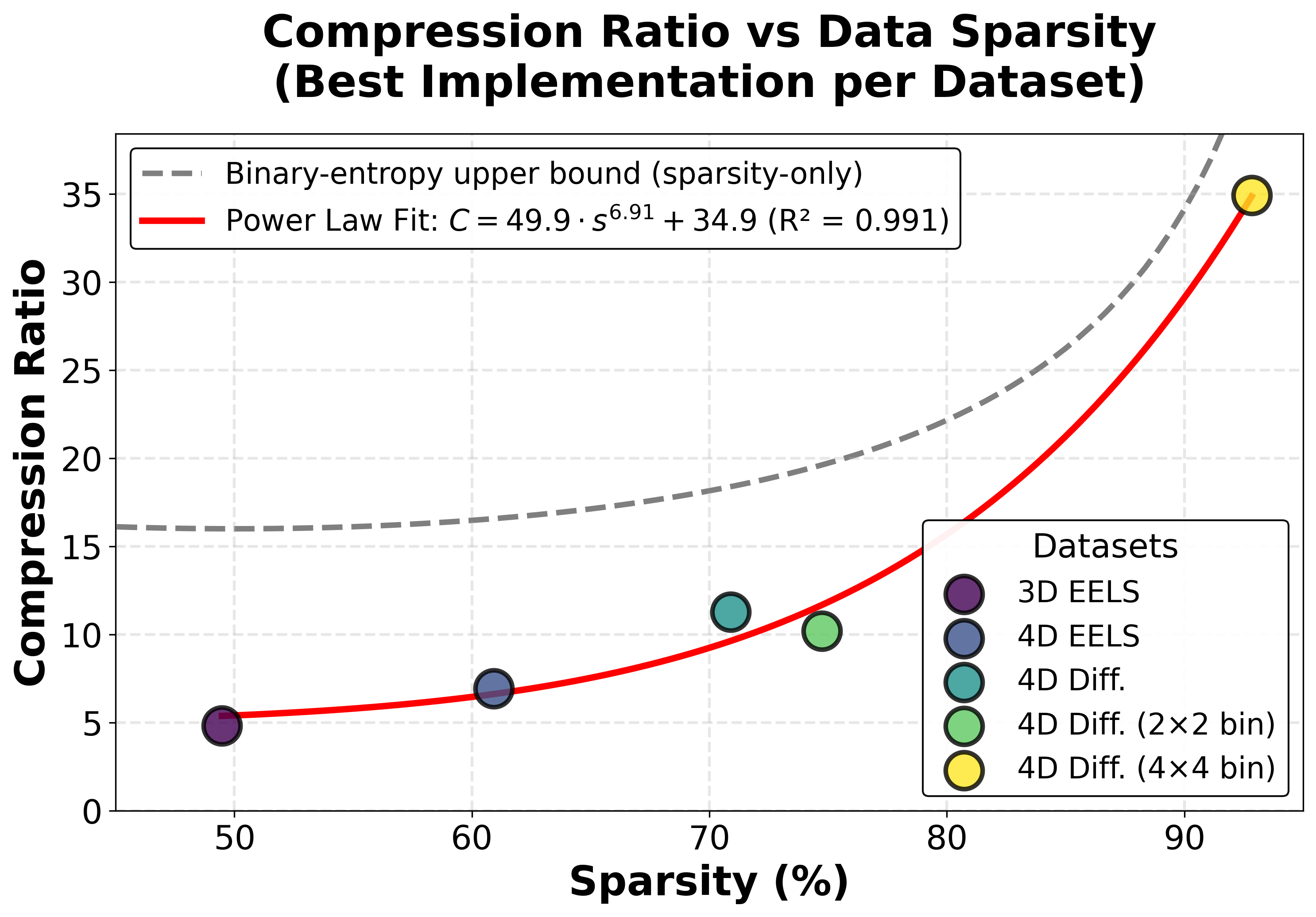}
\caption{Compression ratio as a function of data sparsity across five datasets. Each point represents the best compression achieved for a dataset (using Blosc Zstd or Blosc Zlib). The solid red line shows the power-law fit: $C = 50.0 \times s^{6.90} + 5.0$ ($R^2 = 0.991$). The dashed black line shows a sparsity-only binary-entropy upper bound ($16/H_2(s)$) based on the zero/non-zero indicator. The non-linear relationship demonstrates that compression benefits increase dramatically at high sparsity levels, with highly sparse datasets ($>$90\% zeros) achieving compression ratios exceeding 30$\times$.}
\label{fig:sparsity_compression}
\end{figure}

Compression performance exhibited a strong non-linear relationship with data sparsity (Figure~\ref{fig:sparsity_compression}). Across the five datasets, compression ratios scaled approximately as a power law:

\begin{equation}
C = 50.0 \times s^{6.90} + 5.0
\end{equation}

\noindent where $C$ is the compression ratio and $s$ is the sparsity (fraction of zeros). This model fits the empirical data with $R^2 = 0.991$, indicating that sparsity is the dominant factor determining compressibility.

The non-linear relationship can be understood from information-theoretic considerations. Let $p_0$ denote the probability that a pixel value is exactly zero (i.e., the sparsity, $p_0=s$). The Shannon entropy~\cite{shannonMathematicalTheoryCommunication1948} of the corresponding binary zero/non-zero indicator is the binary entropy function:

\begin{equation}
H_2(p_0) = -p_0 \log_2(p_0) - (1-p_0) \log_2(1-p_0)
\end{equation}

For data stored as 16-bit integers (\texttt{uint16}), a sparsity-only upper bound on the achievable compression ratio is $16/H_2(p_0)$ (this ignores the additional entropy carried by the non-zero values and therefore represents an optimistic limit). At 50\% sparsity, $H_2 \approx 1.0$ bit, yielding an upper bound of 16$\times$. At 93\% sparsity, $H_2 \approx 0.37$ bit, yielding an upper bound of 43$\times$. Our empirical results approach this sparsity-only bound for highly sparse data (34.9$\times$ observed vs 43$\times$ bound).

The fitted exponent (6.90) should be interpreted as a dataset-specific parameter rather than a universal constant, since our benchmark includes heterogeneous acquisition modes, value distributions, and storage dtypes (\texttt{uint16} and \texttt{float32}). The central result is the strong non-linear trend: compression benefits increase dramatically at high sparsity. A dataset with 90\% sparsity achieves approximately 29$\times$ compression, while 50\% sparsity yields only 5$\times$---a 6-fold difference for a 40 percentage point change in sparsity. This has important implications for experimental design: acquisition parameters that increase sparsity (e.g., lower dose, smaller convergence angle) provide disproportionate storage benefits.

Blosc-zstd maintains superior compression across all sparsity levels, while the gap between Blosc and gzip narrows at lower sparsity. This suggests that Blosc's byte-shuffling and blocking optimizations are most effective for highly sparse data.

\subsection{Chunking Strategy Analysis}

Chunking strategy had minimal impact on compression ratio (typically $<$5\% variation) and modest effects on read/write throughput. Table~\ref{tab:chunking_summary} summarizes the average performance by chunking strategy for the 8~GiB 4D Diff. dataset, averaged across all 13 compression implementations.

\input{Tables/table_chunking_summary.tex}

The compression ratio differences were negligible (6.67--6.93$\times$, $<$4\% variation). Write and read throughput showed modest variation (write: 0.63--0.74~GiB/s, $\sim$17\% range; read: 1.19--1.68~GiB/s, $\sim$40\% range), but these effects were implementation-specific with large variation within each chunking strategy (standard deviation 0.7--1.2~GiB/s for read). This indicates that chunking optimization provides limited benefit and would need to be tailored to specific compression implementations.

\begin{figure}[htbp]
\centering
\includegraphics[width=\textwidth]{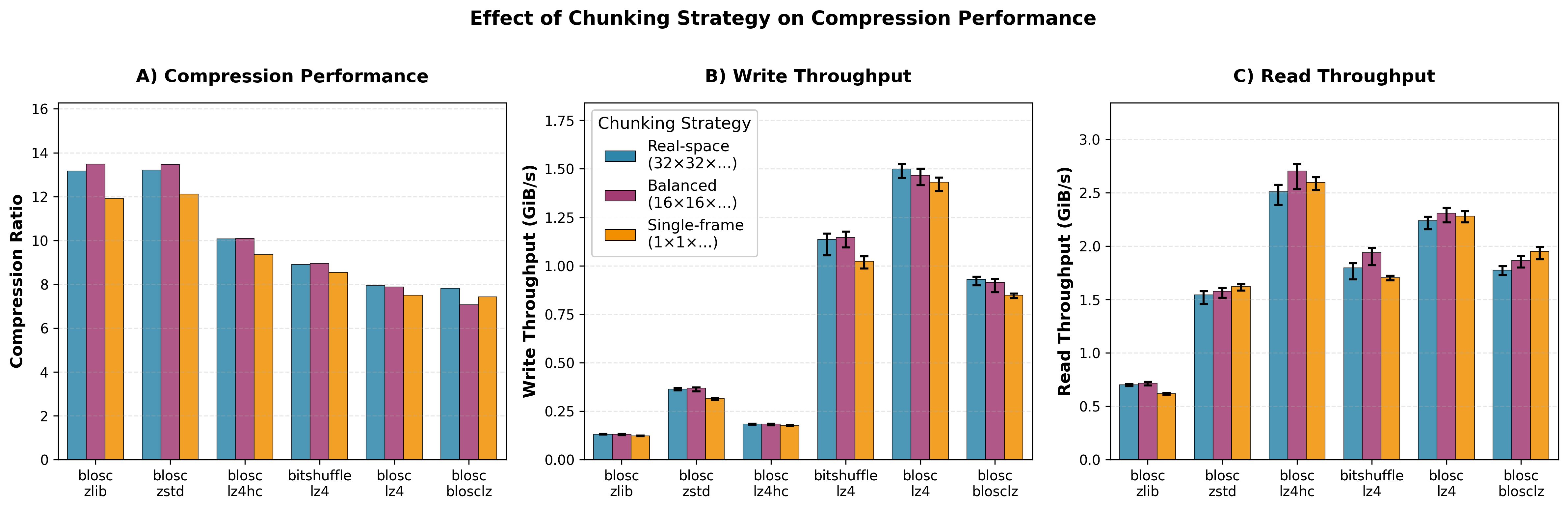}
\caption{Effect of chunking strategy on compression performance for representative implementations. (A) Compression ratios show minimal variation across chunking strategies ($<$5\% difference). Panel A has no error bars as compression ratio is deterministic (0\% variation across 10 runs). (B) Write throughput shows modest variation across strategies, with implementation-specific effects. (C) Read throughput shows modest variation across strategies, with implementation-specific effects. The variation within each implementation across chunking strategies is comparable to the variation between implementations, indicating that chunking effects are implementation-dependent. Error bars on panels B and C show minimum and maximum values across 10 benchmark runs. Six key implementations are shown (see Figure 2 caption for selection criteria). Three chunking strategies are compared: real-space (32$\times$32$\times$256$\times$256), balanced (16$\times$16$\times$128$\times$128), and single-frame (1$\times$1$\times$256$\times$1024). Results averaged across all five datasets.}
\label{fig:chunking_comparison}
\end{figure}

Figure~\ref{fig:chunking_comparison} shows the trade-offs between chunking strategies across the six key implementations. Compression ratios remain relatively constant across chunking strategies (typically within 5\%). Read and write throughput show modest variation with chunking strategy, with implementation-specific effects. The variation within each implementation across chunking strategies is comparable to or smaller than the variation between implementations, indicating that chunking effects are secondary to the choice of compression implementation itself.

\subsection{Sparse Matrix and Custom Compression Methods}

Sparse matrix storage (CSR format) and ad hoc data-reduction strategies did not outperform standard HDF5 compression implementations. Across the five datasets, CSR storage provided modest and inconsistent size reduction (0.63$\times$ to 4.60$\times$), reflecting both index/metadata overhead and the lack of large contiguous zero blocks in typical diffraction patterns~\cite{saadIterativeMethodsSparse2003}. Likewise, the uint8+overflow strategy yielded limited and dataset-dependent compression (1.53$\times$ to 4.00$\times$) and remained substantially worse than Blosc-based methods on the highly sparse datasets.

Two illustrative examples highlight this behavior. For the 2.0~GiB 4D\_EELS dataset (92.8\% zeros), CSR achieved 4.60$\times$ compression (445~MiB), whereas balanced\_blosc\_zstd achieved 34.92$\times$ (58.6~MiB), producing a 7.6$\times$ smaller file. For the 8.0~GiB 4D\_Diff dataset (74.7\% zeros), CSR achieved 1.30$\times$ compression (6.28~GiB), while real\_space\_blosc\_zstd achieved 10.20$\times$ (803~MiB), again yielding a 7.8$\times$ smaller file.

Overall, these results indicate that practical storage and I/O benefits for 4D-STEM datasets are best achieved using established compression implementations.

\subsection{Binning Effect on Compression}

Binning the diffraction patterns reduces uncompressed file size and can reduce sparsity, which may reduce compressibility for aggressive binning. For the 4D Diff. dataset series:

\begin{itemize}
\item No binning (256$\times$256): 8.0~GiB, 74.7\% sparse, 10.2$\times$ compression $\rightarrow$ 803~MiB (real\_space\_blosc\_zstd)
\item 2$\times$2 binning (128$\times$128): 4.0~GiB, 70.9\% sparse, 11.3$\times$ compression $\rightarrow$ 363~MiB (real\_space\_blosc\_zlib)
\item 4$\times$4 binning (64$\times$64): 1.0~GiB, 60.9\% sparse, 6.9$\times$ compression $\rightarrow$ 148~MiB (real\_space\_blosc\_zlib)
\end{itemize}

Binning reduces the final compressed file size, but it also discards information. In this dataset series, the 4$\times$4 binned data compresses to 148~MiB, substantially smaller than the unbinned compressed data (803~MiB), but with reduced detector resolution. In our workflow, binning was performed by averaging and the resulting binned datasets were stored as \texttt{float32}, which changes bytes-per-pixel and the distribution of exact zeros relative to the original \texttt{uint16} data.

It is important to distinguish \emph{compressibility} (compression ratio) from \emph{final storage cost} (final file size). Binning reduces the uncompressed data volume, which generally decreases the final compressed file size even if the compression ratio does not improve. At the same time, averaging can reduce sparsity and alter the value distribution, which can reduce compressibility for aggressive binning. In our 4D Diff. dataset series, sparsity decreases from 74.7\% to 60.9\% between the unbinned and 4$\times$4 binned data, and the best compression ratio decreases from 10.2$\times$ to 6.9$\times$. Nevertheless, combining binning with lossless compression yields the smallest absolute files. Here, we convert an 8.0~GiB dataset into a 148~MiB file through 4$\times$4 binning and lossless compression (real\_space\_blosc\_zlib), a total reduction of $\sim$54$\times$ in storage cost. This illustrates how lossy reduction (binning) and lossless compression can be combined to achieve order-of-magnitude storage savings when scientifically acceptable. For example, binning or cropping diffraction data may be adequate for producing conventional virtual bright-field (BF) or annular dark-field (ADF) images, but it may be undesirable for applications that demand maximum spectral or angular fidelity, such as high energy-resolution EELS or high-resolution ptychographic reconstruction.

%% file: Tables/table_dataset_summary.tex
\begin{table}[h]
\centering
\caption{Dataset characteristics and best compression performance achieved. Chunking strategy is specified in parentheses where it differs from balanced.}
\label{tab:dataset_summary}
\begin{tabular}{lrrrll}
\hline
Dataset & Size (GiB) & Sparsity (\%) & Best Ratio & Implementation & File Size (MiB) \\
\hline
4D\_EELS & 2.0 & 92.8 & 34.9$\times$ & blosc\_zstd & 58.6 \\
4D\_Diff & 8.0 & 74.7 & 10.2$\times$ & blosc\_zstd (real-space) & 803.2 \\
4D\_Diff-2x2-binning & 4.0 & 70.9 & 11.3$\times$ & blosc\_zlib (real-space) & 363.2 \\
4D\_Diff-4x4-binning & 1.0 & 60.9 & 6.9$\times$ & blosc\_zlib (real-space) & 147.6 \\
3D\_EELS & 0.008 & 49.5 & 4.8$\times$ & blosc\_zlib & 1.7 \\
\hline
\end{tabular}
\end{table}

%% file: Tables/table_implementation_families.tex
\begin{table}[h]
\centering
\caption{Implementation family performance comparison (balanced chunking, averaged across all 5 datasets)}
\label{tab:algorithm_families}
\begin{tabular}{lrrr}
\hline
Implementation Family & Mean Ratio & Range & Mean Write (s) \\
\hline
Blosc (zlib) & 13.5$\times$ & 4.8--34.8$\times$ & 19.3 \\
Blosc (zstd) & 13.5$\times$ & 4.8--34.9$\times$ & 8.3 \\
Gzip-9 & 12.3$\times$ & 4.2--32.6$\times$ & 249.9 \\
Gzip-6 & 11.6$\times$ & 4.1--30.6$\times$ & 47.4 \\
Blosc (lz4hc) & 10.1$\times$ & 3.7--25.2$\times$ & 14.8 \\
Szip & 9.2$\times$ & 2.7--24.7$\times$ & 9.8 \\
Blosc (lz4) & 7.9$\times$ & 3.0--18.7$\times$ & 1.9 \\
\hline
\end{tabular}
\end{table}

%% file: Tables/table_chunking_summary.tex
\begin{table}[h]
\centering
\caption{Chunking strategy impact on performance (8 GiB 4D Diff. dataset, averaged across all implementations)}
\label{tab:chunking_summary}
\begin{tabular}{lrrrr}
\hline
Strategy & Chunk Size & Mean Ratio & Mean Write (GiB/s) & Mean Read (GiB/s) \\
\hline
Real-space & (32, 32, 256, 256) & 6.93$\times$ & 0.73 & 1.67 \\
Balanced & (16, 16, 128, 128) & 6.67$\times$ & 0.63 & 1.19 \\
Single-frame & (1, 1, 256, 256) & 6.88$\times$ & 0.74 & 1.68 \\
\hline
\end{tabular}
\end{table}

%% file: sections/discussion.tex

\subsection{Practical guidance for lossless compression}
The results of this work provide immediate, practical guidance for managing 4D-STEM datasets with standard, widely available software. Across the five datasets considered here, Blosc-based implementations consistently occupy the favorable region of the compression--speed trade space, offering compression ratios comparable to or better than HDF5 built-in gzip while providing substantially higher write and read throughput. Within the Blosc family, different implementations provide distinct trade-offs. Blosc Zlib achieves the highest compression ratios, Blosc Zstd provides nearly the same compression with markedly faster write and read performance, and Blosc LZ4 offers the highest throughput at the cost of reduced compression. These trade-offs suggest that implementation choice should be guided by the dominant workflow constraint (storage footprint versus time-to-first-analysis versus sustained throughput).

Chunking strategy plays a secondary role compared to the choice of compression implementation. While chunk geometry can be tuned for specific access patterns and may yield modest gains for particular implementations, the aggregate effects on compression ratio and throughput are small relative to the differences among compression implementations. In most settings, a balanced chunking strategy provides a reasonable default. Further tuning may be advantageous when (i) a fixed access pattern dominates and (ii) a specific compression implementation is already selected.

\subsection{Lossless compression is insufficient}

Although lossless compression substantially reduces file sizes, it is unlikely to resolve the broader data-management challenge posed by rapidly increasing detector throughput. Contemporary pixelated detectors operate at frame rates approaching hundreds of kiloframes per second in electron microscopy workflows, and high-throughput pixel-array detectors in photon science routinely generate data streams on the order of tens of GB\,s$^{-1}$. The Timepix4 readout architecture, for example, supports multi-link high-speed output and motivates system designs targeting tens of Gb\,s$^{-1}$ per chip \cite{llopartTimepix4LargeArea2022,correaTEMPUSTimepix4basedSystem2024}. Likewise, kilohertz pixel-array X-ray detectors can stream at tens of GB\,s$^{-1}$ and require hardware-accelerated acquisition pipelines to sustain such rates \cite{leonarskiJungfraujochHardwareacceleratedDataacquisition2023}. Commercial electron detectors similarly emphasize high-frame-rate operation and associated data handling constraints \cite{DECTRISARINAHybridpixel,DirectElectronDetection}. In this context, improvements in lossless compression can reduce (but cannot eliminate) the mismatch between acquisition bandwidth and practical storage and I/O capacity.

\subsection{Measurement, representation, and opportunity cost}

At root, the data problem is not merely one of storing arrays efficiently. Detector outputs are best understood as measurements: sensor-mediated evidence from which we aim to infer properties of an underlying physical process. In microscopy practice, the term ``raw data'' usually refers to the earliest 
\emph{stored} artifact (e.g., a 4D datacube), but the detector does not begin with an array of integers. The physical signal in the detector is continuous in time and involves intermediate analog stages (charge generation, amplification, shaping, and readout). A measurement is the set of quantities that the instrument ultimately reports after these stages. The report is made meaningful by an instrument model (e.g., calibration, bandwidth, noise, and thresholding assumptions) that connects the measured values to physical quantities and defines their domain of validity.

Measurement theory makes this reduction explicit: a measurement establishes a structure-preserving mapping (a homomorphism) from empirical relations among objects or events to numerical relations, and this mapping is generally many-to-one rather than an isomorphism \cite{krantzFoundationsMeasurementVolume1971}. Distinct objects may therefore share the same numerical description under a given measurement, because the measurement retains only the attribute(s) relevant to the intended inference. For example, a length measurement can be sufficient for inference about geometric fit while discarding other attributes; conversely, for inference about mechanical strength, composition and microstructure may be essential and length may be secondary.

Viewed this way, measurement already discards information in a concrete and familiar sequence of steps. First, integration over a finite exposure time collapses the full time history of the sensor response into a single value per pixel per frame. Second, digitization quantizes that analog value into discrete levels (e.g., a 16-bit integer), discarding variation smaller than one quantization step. Third, common rate-reduction operations such as windowing, binning, and bit-depth selection further reduce the number of reported samples while still producing the same kind of object (smaller frames or smaller datacubes). These operations do not attempt to identify ``objects'' in the data; they reduce volume by sampling the signal more coarsely.

At higher levels of reduction, instruments increasingly return interpreted quantities rather than minimally processed samples. A lock-in amplifier provides an instructive analogy: instead of storing the full waveform, it isolates a narrow frequency component assumed to contain the signal of interest and reports that component while discarding the rest. Event-driven detectors implement an even stronger form of \emph{model-based reduction}. Rather than streaming a dense frame of outputs from every pixel, the detector (or upstream acquisition system) identifies an ``event'' and records only a small set of attributes such as time of arrival, time over threshold, and $(x,y)$ coordinates. In doing so, it discards the underlying sensor response that led to the event classification; the stored object is the interpretation (an event with attributes), not the full set of samples that could in principle have been recorded.

This epistemic framing clarifies why irreversible reduction is not inherently unscientific or poor practice. The relevant question is not whether information is discarded---information is always discarded somewhere along the measurement chain---but whether the retained representation is sufficient for the intended inference (i.e., an \emph{inference-sufficient} representation).

Compressive sensing provides a further example of inference-oriented reduction: rather than recording a fully sampled signal, acquisition is designed so that a reduced set of measurements is sufficient to recover a target representation under explicit structural assumptions (e.g., sparsity in a chosen basis) \cite{candesRobustUncertaintyPrinciples2006,donohoCompressedSensing2006,foucartMathematicalIntroductionCompressive2013,becheDevelopmentFastElectromagnetic2016}. In this view, the goal is not to compress a fully acquired datacube, but to avoid acquiring measurements that are not needed for the intended inference. In particular, lossless compression preserves the stored measurement trace at a certain point along the path to interpretation, which is valuable for reproducibility and re-analysis, but focusing exclusively on lossless compression can obscure the scientific purpose of acquisition: converting measurements into reliable inferences about the physical process.

The increasing use of model-based reduction also highlights an unavoidable opportunity cost in acquisition and storage. Storage capacity, bandwidth, and downstream I/O budgets are finite; allocating them to retain maximal detail limits the total volume of experiments that can be collected, retained, shared, and reprocessed. The traditional argument for preserving all raw data---that it may enable unanticipated future analyses---implicitly assumes storage and I/O is effectively limitless. In high-throughput 4D-STEM regimes (and ``big data'' in general), the relevant decision is therefore not only \emph{how} to compress data, but \emph{which representation} to retain so that the stored record remains sufficient for the intended inference.

A practical implication is that representation should be treated as an experimental design variable. In some contexts, retaining fully dense, losslessly compressed data will remain appropriate. In other contexts, especially those that prioritize rapid feedback or sustained acquisition, it may be preferable to store reduced representations that preserve task-relevant information while discarding that which is deemed to be irrelevant. Event-based representations provide one concrete example of this principle.

To help make this idea operational, we summarize a simple framework for selecting and evaluating inference-sufficient representations. The intent is not to prescribe a single ``correct'' representation, but to clarify what must be specified before any irreversible reduction can be justified.

\begin{quote}
\small\textit{Design principles for inference-sufficient representations:}
\begin{enumerate}
    \item \textit{Specify the inference target(s):} What decision, estimate, or hypothesis test must the stored record support?
    \item \textit{Specify the tolerable error:} What level of error would change the scientific conclusion?
    \item \textit{Identify the information that must be preserved:} Determine which invariants, features, or summary statistics are necessary to support the inference within the required uncertainty.
    \item \textit{Choose the least-burdensome representation that preserves them:} Fully dense measurement traces are one option, but not a requirement.
    \item \textit{Preserve provenance sufficient for trust and reproducibility:} Record calibration, processing assumptions, and metadata needed to reproduce the inference.
\end{enumerate}
\end{quote}

\subsection{Outlook}

Taken together, these results support a two-level approach to data management. Lossless compression provides an important baseline for reducing storage and improving I/O without sacrificing the raw measurement trace, and the present benchmark offers practical guidance for selecting among widely available implementations. However, the long-term sustainability of high-throughput measurement will increasingly depend on combining lossless compression with deliberate choices of acquisition parameters and data representations, including real-time or near-real-time interpretation pipelines that store task-relevant summaries rather than fully dense raw measurements. More broadly, this points toward a richer representation landscape---from event streams to task-specific summaries and learned descriptors---selected to preserve the grounds for an interpretation. In throughput-limited regimes, pipelines may store either the grounds for an interpretation or, when appropriate, the interpretation itself.

%% file: sections/conclusions.tex
We benchmarked 13 off-the-shelf compression implementations for 4D-STEM datasets spanning 8\,MiB to 8\,GiB and a wide range of sparsity, evaluating compression ratio, write throughput, read throughput, and reproducibility across 10 runs. Compression ratios varied strongly with dataset characteristics, with the highest ratios achieved for the most sparse datasets, establishing a reproducible baseline.

Across all datasets, Blosc-based implementations provided the best overall balance of compression and performance. Blosc Zlib achieved the highest compression ratios, whereas Blosc Zstd delivered nearly the same compression with substantially faster write and read performance. For throughput-limited scenarios, Blosc LZ4 offered the fastest write performance and among the fastest read performance at the cost of reduced compression.

Finally, as detector data rates continue to increase, lossless compression alone will not eliminate storage and I/O constraints. Lossless compression preserves a measurement trace at a chosen point in the acquisition chain, but sustainable high-throughput acquisition increasingly requires selecting representations that are sufficient for the intended inference. In this view, compression is a valuable baseline. However, longer-term scalability will depend on combining efficient lossless compression with deliberate representation choices (including model-based reduction where appropriate) to maximize scientific value per stored byte.